\documentclass[runningheads]{llncs}
\usepackage[T1]{fontenc}
\usepackage{graphicx}
\usepackage{algorithm}
\usepackage{algorithmic}
\usepackage{newfloat}
\usepackage{listings}
\usepackage{amsmath}
\usepackage{amssymb}
\usepackage{multicol}
\usepackage{stmaryrd}
\authorrunning{H. Jiang et al.}

\newcommand{\solver}{$\mathsf{HornStr}$}
\usepackage{siunitx}
 \usepackage[flushleft]{threeparttable}
\usepackage{booktabs} 
\usepackage{multirow} 
\usepackage{pifont}
\newcommand{\OMIT}[1]{}

\usepackage[backgroundcolor=orange!50, textsize=scriptsize,textwidth=1cm]{todonotes}

\newcommand{\lang}{\mathcal{L}}
\usepackage{orcidlink}  

\usepackage{tikz}
\usetikzlibrary{shapes.geometric, arrows, positioning,calc, backgrounds}

\tikzstyle{process} = [rectangle, minimum width=3cm, minimum height=1cm, text centered, draw=black]
\tikzstyle{box} = [rectangle, minimum width=1.5cm, minimum height=1cm, text centered, draw=black, rounded corners]
\tikzstyle{arrow} = [thick,->,>=stealth]
\begin{document}

	\title{\solver{}: Invariant Synthesis for Regular Model Checking as Constrained Horn Clauses (Technical Report)}

\author{Hongjian Jiang\inst{1}\orcidlink{0009-0006-4082-2633} \and Anthony W. Lin\inst{1,2}\orcidlink{0000-0003-4715-5096} 
	\and 
	Oliver Markgraf\inst{1}\orcidlink{0000-0003-4817-4563}  \and  Philipp Rümmer \inst{3,4}\orcidlink{0000-0002-2733-7098}  \and Daniel Stan\inst{5,6}\orcidlink{0000-0002-4723-5742}}

\institute{University of Kaiserslautern-Landau, Kaiserslautern Germany,\\
	\and
	Max Planck Institute for Software Systems, Kaiserslautern, Germany \\ 	
	\and 
	University of Regensburg, Regensburg, Germany \\
	\and 
	Uppsala University, Uppsala, Sweden \\
	\and
    EPITA, Laboratoire de Recherche de l’EPITA (LRE), 14-16 Rue Voltaire, 94270 Le Kremlin Bicêtre, France
    \and
    Université de Strasbourg, CNRS, ICube UMR7357, F-67000 Strasbourg, France
    }

\titlerunning{\solver{}: A string Theory Solver for Constrained Horn Clauses }

\maketitle

\begin{abstract}
	We present \solver{}, the first solver for invariant synthesis for Regular
	Model Checking (RMC) with the specification provided in the SMT-LIB 2.6 theory of 
	strings. It is well-known that invariant synthesis for RMC subsumes various 
	important verification problems, including safety verification for 
	parameterized systems. To achieve a simple and standardized file format, 
	we treat the invariant synthesis problem as a
	problem of solving Constrained Horn Clauses (CHCs) over strings.
	Two strategies for synthesizing invariants in terms of regular 
	constraints are supported: (1)~L* automata learning, and (2)~SAT-based 
	automata learning. \solver{} implements these strategies
	with the help of existing SMT solvers for strings, which are
	interfaced through SMT-LIB. 
	\solver{} provides an easy-to-use interface for string solver
	developers to apply their techniques to verification. At the
        same time, it allows verification researchers to painlessly tap into the wealth of 
	modern string solving techniques.
	To assess the effectiveness of \solver{}, we conducted a comprehensive
	evaluation using benchmarks derived from applications including
	parameterized verification and string rewriting tasks. Our experiments
	highlight \solver{}'s capacity to effectively handle these benchmarks,
	e.g., as the first solver to verify the challenging MU puzzle 
	automatically. Finally, \solver{} can be used to automatically
	generate a new class of interesting SMT-LIB 2.6 string constraint 
	benchmarks, which might in the future be used in the SMT-COMP strings track.
	In particular, our experiments on the above 
	invariant synthesis benchmarks produce more than 30000 new \texttt{QF\_S}
	constraints. 
	We also detail the performance of various integrated string 
	solvers, providing insights into their effectiveness on our new benchmarks.
\end{abstract}

\section{Introduction}

\paragraph{Regular Model Checking (RMC)}
\cite{Parosh12,WB98,KMMPS01,RMC,rmc-survey-21} is a prominent framework for modeling an infinite-state transition system as a string rewrite system. Classically, the transition relation is specified as a length-preserving transducer. It is well-known that RMC can be used to model a variety of systems, most notably \emph{parameterized systems}, i.e., distributed protocols with an arbitrary number of processes. Many RMC tools have been developed, focusing on safety verification, e.g., \cite{Lever,AHH13,abdulla2007regular,RMC,WB98,BHRV12,Neider13,Parosh12,TORMC,ChenHLR17,markgraf2020parameterized}, to name a few.

Despite the amount of work on RMC in the past decades and the potential of RMC
in addressing highly impactful verification problems, RMC tools are typically
cumbersome to use. The first problem is the need for the user to specify the 
model in a
low-level language, usually in terms of transducers. The second problem is the
absence of a standard file format agreed upon by RMC tool developers.
Perhaps this is one main reason that most RMC tools attracted very few users
and are mostly no longer maintained today.

\textit{SMT-LIB 2.6 Theory of Strings.}  
String constraints have been standardized as part of SMT-LIB 2.6 since 2020, enabling the organization of a track for string solvers at the annual SMT-COMP.  
The theory over strings has since attracted significant interest in academia
\cite{DBLP:journals/pacmpl/ChenFHHHKLRW22,gutierrez1998solving,10.1145/2743014}
and industry \cite{Neha22,string-sat,Cook18}. The theory provides rich support
for string operators (concatenation, replace-all, regular constraints, length
constraints, etc.), allowing one to conveniently express operations performed in
string-manipulating programs in a high-level language like JavaScript. Out of
the many existing string solvers
\cite{abdulla2015norn,mora2021z3str4,kiezun2009hampi,saxena2010symbolic,tateishi2013path,yu2010stranger,s3p,certistr}, at least five solvers (Z3 \cite{z3}, Z3-alpha \cite{z3-alpha}, Z3-noodler \cite{z3-noodler}, cvc5 \cite{cvc5}, and OSTRICH \cite{ostrich}) now support the SMT-LIB 2.6 format.

\textit{RMC meets String Solvers.}  
In this paper, we propose to connect RMC with string solvers. Our goal is to
provide an easy-to-use and \emph{unified} interface: (i)~for string solver
developers to apply their techniques to verification, and (ii)~for verification
researchers/users who could benefit from RMC and parameterized verification to
easily tap into the wealth of modern string-solving techniques. To this end, we propose to \emph{treat invariant synthesis for RMC as a sub-problem of
	Constrained Horn Clauses (CHCs) over the theory of strings.} CHCs
\cite{bjorner2015horn,grebenshchikov2012synthesizing} form a fragment of
first-order logic over background theories that serves as an intermediate
language for expressing safety verification problems. A CHC
formulation of RMC benefits from the \emph{standard and familiar SMT-LIB specification
	language}. Before our work, no existing CHC solvers directly supported the 
theory of strings.

Our \textbf{first contribution} is, therefore, to develop the first solver \solver{} for invariant synthesis for RMC expressed as a CHC problem over strings. Our solver \solver{} supports two strategies for synthesizing invariants in terms of regular constraints: (1)~L* automata learning \cite{angluin1987learning}, and (2)~SAT-based automata learning \cite{heule2010exact}. Both solvers interact with a string solver via \emph{equivalence queries}, which ask the string solver to verify whether an invariant candidate is correct. The first strategy also interacts with the string solver via \emph{membership queries}, which check whether a guessed string is contained in all invariants.  
To handle both kinds of queries, \solver{} uses other string solvers as backends
through the SMT-LIB 2.6 interface. Note that similar strategies were already
used in other RMC tools \cite{ChenHLR17,Neider13}, where these queries were
answered by interacting with an \emph{ad-hoc} automata implementation, in contrast to string solvers, which are continuously being improved.  
To assess the effectiveness of \solver{}, we conducted a comprehensive evaluation using benchmarks derived from applications, including parameterized verification and string rewriting tasks, integrating the available string solvers individually and in combination. Our experiments highlight \solver{}'s ability to effectively handle these benchmarks, e.g., as the first solver to verify the challenging MU puzzle  automatically.

As a by-product of our tool development, our \textbf{second contribution} is the generation of a new class of \texttt{QF\_S} constraints, which could be used in future SMT-COMP competitions for string solvers. These constraints differ from most benchmarks currently available in SMT-LIB, as they are derived from an invariant synthesis problem. In contrast, the majority of existing benchmarks stem from symbolic execution (like, e.g., the PyEx family). We have evaluated available string solvers on these benchmarks and report the results in this paper.

\section{Constraint Horn Clauses}
We describe in this section the CHC formalism used as input format by \solver{},
as well as examples of applications, illustrating the relationship with
RMC and string-rewrite systems.

\begin{definition}
	A Constrained Horn Clause (CHC) is a first-order logic formula of the form 
	\[ \forall \mathcal{X}. \varphi \land p_1(T_1) \land \dots \land p_k(T_k)
    \to \psi, \; \; (k \geq 0), \]
  in which the term $\psi$ is either an uninterpreted predicate $h(T)$ or $\bot$, and
  $p_1, \dots, p_k$ are uninterpreted predicates. The set of variables
  $\mathcal X$ contains all variables from $T \cup \bigcup^k_{i=1} T_i$.
  The formula $\varphi$ represents a constraint in the background theory,
  such as linear arithmetic or strings.
\end{definition}

A \emph{CHC system} is a conjunction of constrained Horn clauses.
To solve a CHC system, it is necessary to find interpretations of the
uninterpreted predicates that satisfy all clauses.
We focus in the following on CHC systems over the theory of strings,
with one unary uninterpreted predicate. Finding a valuation for this
predicate $p$ amounts to finding a set of words $w$ for which $p(w)$ holds,
so that all clauses are satisfied.
As a finite representation is needed, \solver{} will focus on regular
language solutions.

\subsection{Regular Model Checking}
\begin{example}
  \label{ex:eqdist}
  Consider the token passing protocol on a ring topology, with two initial tokens,
  red and blue,
  at first and last position respectively, moving synchronously in opposite directions,
  without possibly colliding.
  A configuration can be
  seen as a word over $\Sigma= \{r,b,n\}$ where $n$ denotes the
  absence of a token.
  Assume we are interested in the safety property
  ``the two tokens never reach the other end'', invalidated by
  a word in the language
  $\lang(b \cdot n^* \cdot r)$.
  One can observe that an initial odd distance between the two tokens is a necessary
  and sufficient condition for avoiding these configurations.
\end{example}

Checking the safety of this protocol can therefore be specified
with the following CHC System:
\begin{align*}
  V_{i} \in \lang(rn(nn)^*b) &\rightarrow p(V_{i}) \tag{1} \\
  p(V_i)  \land V_i \in \lang(bn^*r) &\rightarrow \bot \tag{2} \\
  p(V_i) \land V_i = A\cdot (rn) \cdot B \cdot (nb) \cdot C \land
          V_o = A\cdot (nr) \cdot B \cdot (bn) \cdot C &\rightarrow
          p(V_o) \tag{3} \\
p(V_i) \land V_i = A\cdot (nb) \cdot B \cdot (rn) \cdot C \land
          V_o = A\cdot (bn) \cdot B \cdot (nr) \cdot C &\rightarrow
          p(V_o) \tag{4}\\
p(V_i) \land V_i = A\cdot (rb) \cdot B \land
          V_o = A\cdot (br) \cdot B &\rightarrow
          p(V_o) \tag{5}
\end{align*}
The variables~$V_i, V_o, A, B, C$ in the clauses are implicitly universally
quantified.
The clauses can be partitioned into three categories:
\(Init=\{ (1)\}\) expresses membership of an initial configuration, while
\(Bad=\{ (2)\}\) expresses undesired configurations. The rest of the clauses,
\(Tr=\{(3),(4),(5)\}\), model
the different transitions where tokens move synchronously, possibly changing
their order in $(5)$.
Note that arbitrarily many extra string variables may
be used as long as they are universally quantified.
The different constraints involve string constraints either in the form of regular
expression constraints ($(1)$ and $(2)$) or in terms of string equality
with concatenation operations ($(3)-(5)$).

Example~\ref{ex:eqdist} is a rather usual instance of RMC
problem, where one asks whether a system is safe by finding
an \emph{inductive invariant}, that is, a set of states or words containing
all initial states $(1)$, no bad state $(2)$, and that
is closed under the transitions $(3)-(5)$.

Several candidate sets can be
considered, such as the set of all reachable words from an
initial clause (the strongest possible invariant), or the set of words
from which no bad state can be reached (the weakest possible invariant).
Recall, however, that we need to compute finite representations of
the considered invariants; in our case, as regular languages.
The previously mentioned sets are therefore less useful:
any reachable and any unsafe configuration must have tokens at equal
distance for the word borders, making the language irregular.
However, a suitable regular inductive invariant does exist,
for example
$\lang(n^*\Sigma(n(nn)^*)\Sigma n^*)$, which translates to
``an odd distance between two tokens''.

\subsection{String-rewrite system: The MU puzzle}
The previous CHC system provided an example of a Regular Model Checking problem
for a system with an initial state of arbitrary length, but where
transitions preserve the length of the word.
Such transitions can usually be represented by length-preserving transducers.
\solver's input formalism is, however, not restricted to
this setting, and can, for example, be applied to string-rewrite systems:

\begin{example}
	\label{ex:mupuzzle}
		The MU puzzle~\cite{10.5555/520739} is a string-rewrite system
	over the alphabet $\Sigma = \{M,I,U\}$:
	Its objective is to determine whether the string $MU$ can be derived from the
	Initial string $MI$ by applying the given rewriting rules:
	$R =\{ (xI \to xIU), (Mx \to Mxx), (xIIIy \to xUy), (xUUy \to xy) \mid x, y \in \Sigma^* \}.$
	For example, using the first rule, the string $MI$
	is transformed to $MIU$ in one step.
\end{example}

We can model the puzzle using the following CHCs, where all variables
$V_i, V_o, x, y \in \Sigma^*$ are universally quantified:
\begin{align*}
V_{i} = MI &\rightarrow p(V_{i}) \tag{1} \\
p(V_{i}) \land V_{i} = xI \land V_{o} = xIU &\rightarrow p(V_{o}) \tag{2} \\
p(V_{i}) \land V_{i} = Mx \land V_{o} = Mxx &\rightarrow p(V_{o}) \tag{3} \\
p(V_{i}) \land V_{i} = xIIIy \land V_{o} = xUy &\rightarrow p(V_{o}) \tag{4} \\
p(V_{i}) \land V_{i} = xUUy \land V_{o} = xy &\rightarrow p(V_{o}) \tag{5} \\
p(V_{i}) \land V_{i} = MU  &\rightarrow \bot \tag{6}
\end{align*}
This CHC system is satisfiable, proving that the MU puzzle cannot
be solved.
%
%

\section{Architecture of \solver{}}
\label{sec:2}
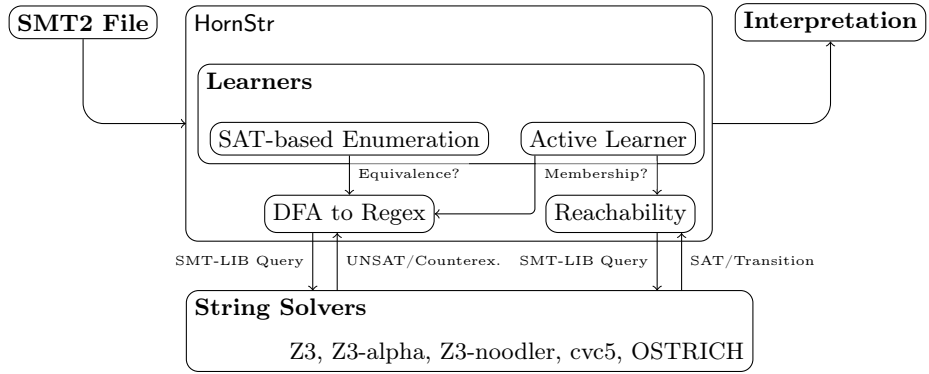
\begin{figure}[tb] \centering 
\begin{center}
\begin{tikzpicture}[
  io/.style={draw, rounded corners=0.5em},
  tool/.style={draw, rounded corners=0.5em},
  boxlabel/.style={font=\bfseries},
]

  \begin{scope}[every node/.style={io}]
    \node[anchor=west] (input) at (0,0) {\small \textbf{SMT2 File}};
    \node[anchor=east] 
        (output)
      at (\textwidth,0)
        {\small \textbf{Interpretation}};
   \end{scope}
  \begin{scope}[local bounding box=chcsolver]
    \node[boxlabel]
            at ($(input.east)!0.1!(output.east)$) (solverlabel) {\solver};
    \begin{scope}[local bounding box=learner]
      \node[boxlabel, anchor=north west, node distance=2em]
        at ($(solverlabel.south west)+(0.5em,-1em)$)
        (learnerlabel)
          {Learners};
      \node[tool, anchor=north west] at ($(learnerlabel.south west)+(0.5em,-1em)$)
          (satbased)
          {SAT-based Enumeration};
      \node[tool, right of=satbased, node distance=7em, anchor=west]
          (activebased) {Active Learner};
      \node (spacer) at ($(activebased.south east)+(0.0em,-0.0em)$) {};
    \end{scope}
    \node (spacer) at ($(learner.east)+(0.0em,-0.0em)$) {};
    \node[below of=satbased, tool] (dfa2regex) {\small DFA to Regex};
    \draw[->, rounded corners=2pt] ($(activebased.south)+(-3em,0)$)
        |- (dfa2regex);
    \node (spacer) at ($(dfa2regex.south)+(0,0em)$) {};
    \draw (satbased) edge[->]
      node[right, fill=white, fill opacity=0.5, text opacity=1] {\tiny Equivalence?} (dfa2regex);

    \node[below of=activebased, tool, xshift=0.5em] (reach) {\small Reachability};
    \draw let \p1=($(activebased.south)+(2em,0)$) in
          let \p2=(reach.north) in
      (\x1,\y1)
      edge[->]
      node[left, fill=white, fill opacity=0.5, text opacity=1] {\tiny
        Membership?} (\x1,\y2);
  \end{scope}
  \begin{scope}[local bounding box=strsolver]
    \node[boxlabel,anchor=north west]
        at ($(chcsolver.south west)+(0,-2em)$) (strsolverlabel) {String Solvers};
    \node[below of=strsolverlabel, anchor=north west, node distance=1em]
              {Z3, Z3-alpha, Z3-noodler, cvc5, OSTRICH};
  \end{scope}
  \begin{scope}[on background layer]
    \draw[io] (chcsolver.north west) rectangle (chcsolver.south east);
    \draw[io] (learner.north west) rectangle (learner.south east);
    \draw[io] (strsolver.north west) rectangle (strsolver.south east);

    \coordinate (start) at (dfa2regex.south);
    \path let \p1 = (strsolver.north) in
          let \p2 = (start) in
        coordinate (end) at (\x2,\y1);
    \draw[transform canvas={xshift=-1.5em}] (start) edge[->]
        node[left] {\tiny SMT-LIB Query}
      (end);
    \draw[transform canvas={xshift=-0.5em}] (end) edge[->]
        node[right] {\tiny UNSAT/Counterex.}
        (start);
    \coordinate (start) at (reach.south);
    \path let \p1 = (strsolver.north) in
          let \p2 = (start) in
        coordinate (end) at (\x2,\y1);
    \draw[transform canvas={xshift=1.5em}] (start) edge[->]
        node[left] {\tiny SMT-LIB Query}
      (end);
    \draw[transform canvas={xshift=2.5em}] (end) edge[->]
        node[right] {\tiny SAT/Transition}
        (start);
  \end{scope}

  \begin{scope}[rounded corners=2ex]
    \draw[->] (input.south) |- (chcsolver.west);
    \draw[->] (chcsolver.east) -| (output);
  \end{scope}
\end{tikzpicture}
\end{center} 
	 \caption{The overall framework of \solver.} 
	 \label{fig: framework} 
\end{figure}

The \solver{} framework integrates CHC and string constraints, leveraging automata learning techniques in combination with string solvers. This integration addresses complex problems expressed in SMT-LIB files, and modeling, e.g., parameterized
systems or string-rewrite systems. Figure \ref{fig: framework} illustrates the overall architecture of \solver{}.

The framework commences with an SMT-LIB formatted file as its input, a format prevalent in the SMT community for describing problems that require solutions to satisfy constraints involving complex data types and operations. The 
\emph{Learners} play a crucial role in synthesizing predicates based on regular constraints. They employ two different strategies: \begin{enumerate} 
\item \emph{SAT-based Enumeration} utilizes SAT solvers to generate potential solutions, as well as string solvers to assess whether the solution satisfies given CHCs. In case of violated CHCs, the string solvers can provide a counterexample. Initially, the set of counterexamples is empty. The learner constructs a Deterministic Finite-state Automaton (DFA) as a hypothesis solution that accepts every word. This DFA is transformed into a regular expression via an intermediate translator, implemented by Brzozowski and McCluskey's state elimination method \cite{brzozowski1963signal}. The translator then sends an SMT-LIB query to the \emph{String Solvers} to check for consistency with the CHCs. Upon receiving the query, the string solvers check the solution behind the scenes, returning either $unsat$ or a counterexample to the learner. 
	\item \emph{Active Learner} directly interacts with the learning model through queries. This learner constructs both equivalence queries and membership queries to verify if a string or sequence belongs to the model's language or reachability queries to determine if a certain state or condition is achievable. It maintains an observation table~\cite{angluin1987learning} in its cache, from which it constructs a DFA. The \emph{Reachability} module is responsible for communicating with the string solvers to ascertain whether the queried word is within the language: this involves several string queries, enumerating initial words ($Init$), then using all applicable transitions ($Tr$) to find all reachable words iteratively.
	The membership query is answered positively when the desired word is found in the reachable fragment, or negatively if all the words of
	the same length, or up to a fixed constant, have been explored. The latter rule constitutes a heuristic inspired by the length-preserving transition models.
\end{enumerate}

Furthermore, String Solvers respond to queries from the Learners by resolving a series of string constraints.
To enhance the efficiency of answering equivalence and membership queries, the framework has integrated an incremental solving technique.
Each CHC is assigned to a dedicated solver thread, for pre-computation purposes. 
For each word or automata query, the system saves the current constraints (push), inserts the new query constraint, computes the result, and upon obtaining the result, it restores the saved constraints (pop), provides the response, and prepares for the next query.
Through a command line argument, the user can also instruct \solver{} to handle all CHCs using a single solver, as this may save processing time for larger equivalence queries.
On the contrary, word queries involve small input values, so they usually beneficit from specific String Solver optimizations, one for each clause.
\solver{} employs a variety of state-of-the-art solvers, such as Z3, Z3-alpha, Z3-noodler, cvc5, and OSTRICH. Each of these solvers brings unique capabilities that range from basic string manipulations to more complex pattern matching and replacement operations. These specialized tools are adept at managing string operations within the constraints specified in SMT-LIB queries. Additionally, the framework offers a configuration file for users to specify their own string solver, as an external implementation of the interactive mode of the SMT-LIB 2.6 standard.

The execution of \solver{} progresses through the following phases:
\begin{enumerate}
	\item \textbf{Initialization}: The procedure begins with the selection of a suitable learning strategy and a string solver. Subsequently, an SMT-LIB file containing the uninterpreted predicate declaration, followed by constraint Horn clauses, is loaded, and the designated solver is instantiated together with the necessary oracles.
	
	\item \textbf{Query Processing and Model Refinement}: When employing the
    SAT-based enumeration approach~\cite{Neider13}, the learner initially constructs a DFA with a single state and an empty
    counterexample set using a SAT solver. Once an appropriate DFA is generated
    that integrates the counterexample set, an equivalence check is conducted
    against the hypothesis using the string solver. If a new counterexample is
    detected, the hypothesis undergoes refinement and reconstruction. If the SAT
    solver returns an unsatisfiable outcome, the automaton's state space is
    incrementally expanded, and the process iterates until the string solver
    fails to find further counterexamples and accepts the hypothesis. 
	
	Alternatively, if the active learner is selected~\cite{ChenHLR17},
    membership queries are
    issued to verify whether a given word \( w \) belongs to the target
    language \( L \), leveraging the reachability module. This initiates an
    iterative process in which membership queries facilitate hypothesis
    generation, which is subsequently validated via equivalence queries.
	
	\item \textbf{Solution Generation}: Based on the preliminary results and constructed queries, the string solver is employed to analyze and resolve regular constraints. Within this framework, the solver integrates the \( Init \) and \( Tr \) components to determine whether a word \( w \) is accepted. Conversely, if the word is rejected, the decision is justified through the \( Bad \) and \( Tr \) components. 
	
	For equivalence queries, all Horn clauses are evaluated by testing them with
    two free variables, \( var_{in} \) and \( var_{out} \). For example, if
    \( var_{out} \) appears as part of a word in the hypothesis and satisfies
    the \( Bad \) clause, it is classified as a negative counterexample.
    Similarly, positive and inductive counterexamples can be identified using
    the \( Init \) and \( Tr \) components, respectively.
    If unsupported by the learner, inductive counterexamples are converted into positive and
    negative counterexamples thanks to reachability analysis, following the
    \textit{strict but generous teacher}~\cite{ChenHLR17} concept.

	
\end{enumerate}

\section{Evaluation}

In this section, we evaluate the performance and capabilities of \solver{}\footnote{\url{https://arg-git.informatik.uni-kl.de/pub/string-chc-lib}} \cite{jiang_2025_15190404} on a set of benchmarks derived from the verification of distributed systems and string rewriting systems. 
\solver{} uses string solvers as oracles for membership and equivalence queries, the choice of the solvers in use is an important aspect of its performance.

Our evaluation is divided into two parts. 
First, we examine how the different string solvers can handle the string formulas generated as queries during the CHC-solving process. 
As described in Section \ref{sec:2}, \solver{} supports incremental solving, which can improve efficiency by reusing information across related queries.
We compare the performance of string solvers on both incremental and non-incremental queries.

Second, we evaluate \solver{}'s overall performance using the string solvers that performed best in the first part of the evaluation.
Experiments were conducted on an Intel Core i7-10510U CPU at 1.8GHz with 16 GB of RAM running on Windows 11.

Our benchmarks are derived from two distinct domains:

\begin{itemize}
	\item \textbf{Verification of Distributed Systems:} 
	We transform Regular Model Checking protocols~\cite{ChenHLR17,esparza_et_al:LIPIcs.CONCUR.2022.23} into Constrained Horn Clause (CHC) programs using automatic translations:  Bakery\cite{lamport2019new}, Szymanski\cite{szymanski1990mutual,gribomont1998automated}, Dijkstra\cite{lynch1996distributed}, Burns\cite{lynch1996distributed}, Dining Philosopher Protocol\cite{hoare1978communicating}, Israeli-Jalfon's self-stabilising protocol\cite{israeli1993uniform}, Resource-allocator protocol\cite{donaldson2007automatic}, David Gries's coffee can problem\cite{lin2016regular}, german protocol\cite{abdulla2007regular} and Kanban production system\cite{geeraerts2006expand}.
	
	\item \textbf{String Rewriting Systems:} 
	We also manually model the MU puzzle and EqDist protocols as CHC programs, demonstrating the versatility of the approach.
\end{itemize}

\subsection{Results of the String Solver Experiments}
Table \ref{tab:compare} provides a comparison of the string solvers Z3, cvc5, Z3-noodler, Z3-alpha, and OSTRICH. 
The benchmarks are categorized into incremental and non-incremental queries, further divided by the query type: membership or equivalence.
Our primary metric of interest is the number of benchmarks solved, as failing to resolve even a single query can prevent the CHC solver from terminating. The timeout for each benchmark is set to 30s.

Equivalence queries predominantly involve reasoning over regular expressions but may also include word equations when these are part of the Horn clause. Membership queries, while also involving regular expressions, tend to emphasize disequalities ($x \neq c$, where $x$ is a string variable and $c$ is a string constant).

In the incremental setting, the membership results are relatively similar, with all solvers processing over 514 benchmarks. Notably, Z3-Noodler leads by solving all 523 benchmarks in an average of 109.7 seconds, whereas OSTRICH, cvc5, Z3, and Z3-alpha solve between 514 and 518 benchmarks in slightly higher runtimes. 

For the incremental equivalence queries, we see different behaviors among the solvers. Z3-noodler solves all 396 queries in just 15.5 seconds, while OSTRICH manages 378. On the other hand, cvc5, Z3, and Z3-alpha only solve between 109 and 126 queries. A similar pattern shows up in the non-incremental equivalence queries: Z3-noodler handles all 848 queries, with OSTRICH coming in close with 784, whereas cvc5, Z3, and Z3-alpha solve between 403 and 457 queries. In the case of membership queries, every solver covers nearly all of the 30,902 benchmarks, with only cvc5 and OSTRICH missing about 1\%, while Z3-noodler and Z3 turn out to be the fastest to solve them all.

Across both incremental and non-incremental benchmarks, the results demonstrate a consistent pattern: membership queries are generally handled well by most solvers, while equivalence queries involving regular expressions remain a challenge for many. 
Notably, automata-based solvers such as Z3-noodler and OSTRICH consistently show superior performance on equivalence queries, likely due to their design being well-suited for reasoning over regular expressions. These results also highlight the high incrementality of our approach, as seen when comparing the total time spent on all incremental vs. non-incremental queries.
Note that OSTRICH's overall runtime is a bit higher partly due to the JVM startup time incurred for each benchmark.

To address the challenges faced by solvers struggling with equivalence queries, we experimented with different settings and flags for those solvers and implemented a regular expression simplifier on our end before sending the queries. The simplifier aimed to reduce the nesting of Kleene stars using algebraic transformations on regular expressions. While this led to marginal improvements for some poorly performing solvers, it had little impact overall and even worsened performance for solvers already handling regular expressions effectively.

\begin{table}[h!]
	\centering
	\caption{Comparison of state-of-the-art string solvers. Benchmarks are divided into incremental and non-incremental membership and equivalence queries. The timeout is 30s. Timeouts are excluded from solved time.}
	\label{tab:compare}
	\begin{threeparttable}
		\begin{tabular*}{\textwidth}{@{\extracolsep{\fill}} l
				S[table-format=3.0] S[table-format=5.1]
				S[table-format=3.0] S[table-format=5.1]
				S[table-format=5.0] S[table-format=5.1]
				S[table-format=3.0] S[table-format=4.1] }
			\toprule
			& \multicolumn{4}{c}{\textbf{Incremental}} & \multicolumn{4}{c}{\textbf{Non-Incremental}}\\
			\cmidrule(lr){2-5} \cmidrule(l){6-9}
			Solver
			& {Mem} & {Time (s)}
			& {Equiv} & {Time (s)}
			& {Mem} & {Time (s)}
			& {Equiv} & {Time (s)} \\
			\midrule
			OSTRICH    & 514 & 453.2 & 378 & 410.8 & 30773 & 16504.9 & 784 & 980.8 \\
			cvc5       & 517 &  97.8 & 126 & 7948.4 & 30652 &   610.5 & 457 & 270.7 \\
			Z3         & 517 &  33.7 & 109 &  506.7 & 30902 &  1511.7 & 403 & 102.1 \\
			Z3-noodler & 523 & 109.7 & 396 &   15.5 & 30902 &   806.3 & 848 &  17.9 \\
			Z3-alpha   & 518 &  86.4 & 109 &  516.6 & 30902 &  3839.1 & 404 & 162.7 \\
			\bottomrule
		\end{tabular*}
	\end{threeparttable}
\end{table}

\subsection{Results of the \textbf{\solver{}} Experiments}

After evaluating the performance of various string solvers as membership and equivalence oracles in our preliminary experiments, we now assess \solver{} for CHC solving. Based on the incremental benchmark results (Table~\ref{tab:compare}), we chose Z3 for membership queries and Z3-noodler for equivalence queries. 

We developed an automatic parser that transforms length-preserving RMC protocols into CHC SMT2 format, incorporating word equations and regular membership constraints. Next, we evaluate the efficiency of \solver{} using both SAT-based Enumeration and the Active Learner, as described in Section~\ref{sec:2}. In our evaluation, we record whether \solver{} produces a deterministic finite automaton for the uninterpreted invariant within a predefined time limit or identifies an unsafe trace during the benchmark evaluation.

Our evaluation demonstrates that our tool solved most benchmarks in under a second using either SAT-based enumeration or the active learner. Notably, SAT-based enumeration solved every protocol listed in Table~\ref{tab: example}, whereas the active learner failed to find solutions for some benchmarks. However, certain protocols—such as \textit{Kanban} and \textit{German}—exceeded the 60-second timeout due to the complexity of transitions in their CHC representations. Detailed evaluation results are presented in Table~\ref{tab: example}.

\begin{table}[h!]
	\centering
\caption{Comparison of protocols: automaton size and learning time across SAT-Based and active Learner }
	\label{tab:learner}
	\begin{tabular}{|c | c  c | c  c | }
		\toprule
		& \multicolumn{2}{c|}{SAT-based Enumeration} & \multicolumn{2}{c|}{Active Learner}  \\
		 Protocol & Size & Time($s$) & Size & Time($s$)\\
		\midrule
		Token Pass & 3 & 0.41 & 3 & 0.10\\
		2 Tokens Pass & 3 & 0.78 &  6 & 0.57\\
		3 Tokens Pass & 2 & 0.30 & 2 & 0.17\\
		Power-Binary & 1 & 0.2 &  1 & 0.01\\
		Bakery &  2 & 0.15 &  3 & 0.37 \\
		Burns & 2 &  2.09 &  \ding{56} & TO \\ 
		Coffee-Can & 2 & 0.52  & 5 & 9.66 \\
		Coffee-Can-v2 & 3 & 0.31  & 4 & 23.45\\
		Herman-Linear & 2 & 0.11 & 2 & 0.08\\
		Herman-Ring & 2 & 0.51 & 2 & 0.33\\
		Israeli-Jalfon & 3 & 0.35 & 4 & 0.46\\
		LR-Philo & 2 & 0.80 & 3 & 2.84\\
		Mux-Array &  2 & 0.49  & \ding{56} &  TO\\
		Resource-Allocator & 2 & 0.14 &  4 & 25.19 \\
		Eqdist & 3 & 1.45 &   \ding{56} &  TO \\
		MU Puzzle & 3 &11.01 &  \ding{56} & TO \\
		Water-Jug & 2 & 2.05 &  \ding{56} & TO \\
		Dining-Crypt & 2 & 10.02 &  \ding{56} & TO \\
		\bottomrule
	\end{tabular}
	\label{tab: example}
\end{table}

\section{Conclusions}

We introduced \solver{}, the first solver for invariant synthesis in RMC that leverages the SMT-LIB 2.6 Theory over Strings. By formulating invariant synthesis as a problem of solving CHCs over strings, \solver{} provides a standardized, scalable, and automated approach to verification. Our approach enables seamless integration of modern SMT solvers into RMC verification, bridging parameterized verification and string solving in a novel way.

Our evaluation demonstrated \solver{}’s effectiveness in handling complex verification tasks, including parameterized systems and string rewriting problems (e.g., the MU puzzle). By integrating incremental solving techniques, \solver{} significantly improves the performance of string solvers, reducing computational overhead and enhancing scalability. Additionally, our work contributes more than 10,000 new \texttt{QF\_S} constraints, providing a valuable benchmark suite for SMT solver evaluations.

We mention several future research avenues. The first is to extend \solver{} 
by handling general CHCs over strings, i.e., non-linear and monadic CHCs that 
permit \emph{symbolic alphabets}. This would allow one to model certain
protocols, wherein process IDs are passed around (e.g. Chang-Roberts protocol;
see \cite{HL24,ivy}).
Second, one could extend our CHC framework to other types of
RMC verification including liveness \cite{LR16,LLMR17} and bisimulation 
\cite{HLMR19,lin2016regular}. 


\clearpage

\bibliographystyle{splncs04}

\bibliography{cav25}



\end{document}